\documentclass[journal]{IEEEtran}
\usepackage{graphicx}

\usepackage{amsmath}
\usepackage{cite}
\usepackage{amsmath,amssymb,amsfonts}
\usepackage{algorithmic}
\usepackage{graphicx}
\usepackage{textcomp}

\ifCLASSINFOpdf
\else
\fi
\begin{document}
%
\title{Multicore fiber-based quantum access network}
%
%
%

\author{Chun~Cai,
        Yongmei~Sun,
        Jianing~Niu,
        Peng~Zhang,
        Yongrui~Zhang,    
        and~Yuefeng~Ji
\thanks{Yongmei Sun is with the The State Key Laboratory of Information Photonics and Optical Communications, Beijing University of Posts and Telecommunications, Haidian District, Beijing, 100876, China
e-mail: ymsun@bupt.edu.cn.}
\thanks{Chun Cai, Yongrui Zhang, Peng Zhang, Jianing Niu, and Yuefeng Ji are with Beijing University of Posts and Telecommunications.}
}

\maketitle

\begin{abstract}
We propose a quantum access network based on multicore fiber (MCF) to scale up the number of users in quantum key distribution (QKD) networks. The MCF is used as feeder fiber and single-core single-mode fibers (SSMFs) are used as drop fibers. Quantum signals (QSs) are integrated with classical signals (CSs) in both MCF and SSMFs to save deployment cost since access networks are cost-sensitive. Due to the integration, spontaneous Raman scattering (SRS) and inter-core crosstalk (IC-XT) are the main impairment sources to QKD. To alleviate the noise, we propose a core and wavelength assignment scheme (CWAS) that a dedicated core of MCF is used to transmit QSs and the wavelengths of QSs are set lower than those of upstream signals. Also, we demonstrate that wavelength-time division multiplexing (W-TDM) is suitable for QSs which are required to support large number of quantum users, since W-TDM can realize higher secure key rate (SKR) than time division multiplexing (TDM) and require lower cost than wavelength division multiplexing (WDM). Finally, the proposed quantum access network is verified experimentally. The experiment results are consistent with our analysis. The properties of SRS in the proposed architecture are shown in the experiments through the quantum bit error rates (QBERs) in different experimental case, which verifies the superiority of the proposed CWAS. Also, the characteristics of the SKRs prove that the number of receivers has a great impact on the performance of QSs using W-TDM.
\end{abstract}


%
\IEEEpeerreviewmaketitle

\section{Introduction}
%
%
%
%
\IEEEPARstart{Q}{uantum} key distribution (QKD) allows remote parties to establish encryption keys by the laws of physics \cite{Bennet1984Quantum,Gisin2001Quantum}. It enables information-theoretic communication security and could revolutionize the way in which information exchange is protected in the future \cite{Shor2000Simple}. Nowadays it has reached the level of maturity required for deployment in real-world scenarios and many practical quantum networks have been proposed \cite{choi2011quantum,azuma2015all,PhysRevLett.120.030501,Mao:18}. An important step in the widespread application of QKD is to expand the number of users in quantum networks. Access networks can be a good solution to this problem due to its point-to-multipoint architecture.

The concept of quantum access networks is first introduced and experimentally demonstrated in \cite{frohlich2013quantum}. Quantum signals (QSs) are transmitted in dedicated fibers in their architecture. All the optical network units (ONUs) are connected to optical line terminal (OLT) with one optical power splitter in their architecture and different QKD transmitters are based on time division multiplexing (TDM). Thus the secure key rate (SKR) will decrease dramatically with the increase of ONU number due to the insertion loss of the splitter and the decrease of QKD transmitting frequency. Then quantum access networks in which QSs are integrated with classical signals (CSs) in the same fiber are proposed \cite{frohlich2015quantum,doi:10.1063/1.5003342}. Noise generated from intense CSs is a great challenge for these schemes, which has been studied widely in point-to-point transmission \cite{sun2016reduction, niu2018optimized}. In \cite{frohlich2015quantum}, QKD is integrated in a gigabit passive optical network (GPON). QSs co-exist with CSs in drop fibers. However, another feeder fiber is required to transmit QS specially due to the spontaneous Raman scattering (SRS) noise from CS, which will increase cost of the network greatly. Also, the SKR will decrease dramatically with the increase of ONU number for the same reason as in \cite{frohlich2013quantum}. In \cite{doi:10.1063/1.5003342}, a bypass structure and an optical switch are used to avoid the insertion loss of the splitter. However, noise still limits the performance of the system.

Optical networks play an increasingly important role in our lives \cite{ji2018towards} and the data traffic demand in access and backbone optical networks has been increased exponentially \cite{yu2013transmission}. Multiplexing over different degrees of freedom in conventional single-core single-mode fiber (SSMF), including wavelength, phase, time and polarization multiplexing, is being utilized to circumvent the future information capacity crunch \cite{essiambre2010capacity}. However, the capacity of existing standard SSMF may no longer satisfy the growing capacity demand and is approaching its fundamental limit around 100 Tbps owing to the limitation of amplifier bandwidth, nonlinear noise, and fiber fuse phenomenon \cite{qian2011101}. In order to further increase the fiber capacity, space division multiplexing (SDM) has been proposed and attracted intensive research efforts as a solution to the capacity saturation of conventional SSMF \cite{saitoh2013multicore,richardson2013space,winzer2014making}. Multicore fiber (MCF) is an effective means to realize SDM. Random power coupling between different cores is called inter-core crosstalk (IC-XT),which is the main factor affecting the performance of MCF. There are two types of MCF, one is weakly coupled multicore fiber (WC-MCF) \cite{hayashi2011design} and the other is strongly coupled multicore fiber (SC-MCF) \cite{arik2013coupled}. WC-MCF has the advantage of lower IC-XT than SC-MCF. Thus WC-MCF is suitable for transmitting QS which is vulnerable to interference and we have studied the impact of IC-XT on QS in WC-MCF \cite{Cai:19}.

The next generation optical access network is required to support large capacity data transmitting to enormous number of users. Many technologies have been proposed to realize it \cite{BINDHAIQ201553, 7402264, Wei:16}. SDM is a promising method to realize the next generation optical access network and several classical access networks based on MCF have been proposed\cite{li2015experimental, feng2017ultra}.

QKD has not been widely implemented in current access networks and we believe that the next generation optical access network is a great opportunity to promote QKD. Thus we propose a quantum access network which can support enormous subscribers. WC-MCF is used as feeder fiber while SSMFs are used as drop fibers. CSs are based on space-wavelength division multiplexing (S-WDM) and can support large amount of users, which is benefit from the rich space and wavelength resources of MCF. To realize QKD in such quantum access networks, cost and SKR are two major problems needed to be considered. In order to save cost, QSs are integrated with CSs in both MCF and SSMFs since we do not need to lay a dedicated fiber for QSs. Also, the expensive QKD receivers are placed in the OLT, which will be shared by several users based on wavelength-time division multiplexing (W-TDM). Noise and channel loss are two main factors for SKR. The main noises in the network are SRS and inter-core crosstalk (IC-XT). In order to alleviate the noises, a core and wavelength assignment scheme (CWAS) is proposed. In MCF, a dedicated core is utilized to transmit QSs to alleviate SRS since SRS has a stronger impact on QSs in the same core. To avoid the impact of IC-XT, different wavebands are assigned to QSs and CSs. Thus IC-XT can be turned to out-of-band noise and will be filtered effectively by bandpass filter. In SSMF, the SRS generated from upstream signals has a greater impact on QSs than that generated from downstream signals and the frequency of upstream signal is set lower than that of QS due to the smaller SRS coefficient. Finally, the proposed quantum access network is verified experimentally. The QBERs under different experiment conditions show that the SRS which generated from upstream signals in SSMF has the greatest impact on QKD in the proposed CWAS. However, the noises can be effectively suppressed with appropriate filtering measures. The SKRs are strongly correlated with the transmitting frequency of each ONU and the insertion loss of the splitter with different splitting ratio. Also, the relationship between the number of receivers and SKRs of each subscribers is shown through the simulation based on our system.

\section{Architecture of the proposed quantum access network}
\label{architecture}

\begin{figure}[h]
\centering\includegraphics[width=8cm]{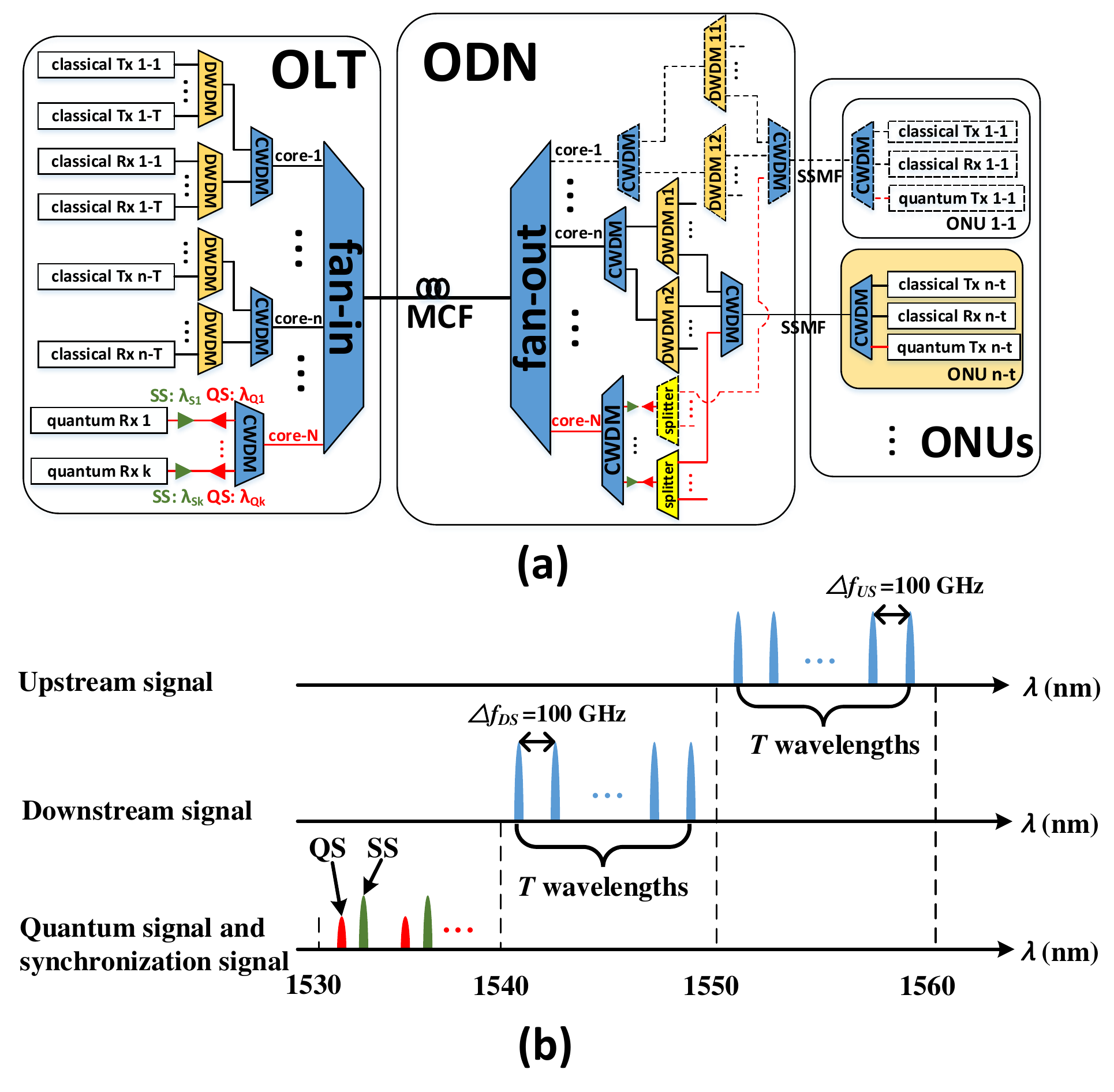}
\caption
{\label{fig:architecture}(a) Proposed MCF-based quantum access network architecture. (SSMF: single-core single-mode fiber, DWDM: dense wavelength division multiplexing module, CWDM: coarse wavelength division multiplexing module, OLT: optical line terminal, ODN: optical distribution network, ONU: optical network unit, MCF: multicore fiber, SS: synchronization signal, QS: quantum signal, Tx: transmitter, Rx: receiver). (b) An example of wavelength assignment.}
\end{figure}

The proposed quantum access network architecture based on a MCF is illustrated in Fig.~\ref{fig:architecture}(a). In the proposed network, MCF is used as feeder fiber and SSMFs are used as drop fibers. QSs are integrated with CSs in both MCF and SSMFs.

For classical communication, $T$ wavelengths are utilized as the downstream channels in the OLT block. They are multiplexed by a dense wavelength division multiplexing (DWDM) module with small frequency spacing. We take Fig.~\ref{fig:architecture}(b) as an example of wavelength assignment in which the downstream channels are put between 1540 and 1550 nm and the frequency spacing is 100 GHz. The $T$ downstream signals are multiplexed with the $T$ upstream signals in the same core by a coarse wavelength division multiplexing (CWDM) module. This is because different wavebands are used for them. The wavelength assignments are the same for the $N-1$ cores which are used to transmit CSs. In optical distribution network (ODN), one CWDM and two DWDM modules are used to demultiplex CSs in the same core. Also a CWDM is utilized to couple upstream signal, downstream signal, QS and synchronization signal (SS) of one ONU into corresponding SSMF. In the ONU block, a CWDM is used to couple different signals into the SSMF. $(N-1)*T$ subscribers can be supported in this configuration.

For QKD, the first issue to consider is the CWAS. IC-XT and SRS are two main factors for CWAS. The power of IC-XT (about -60 dB/km) is higher than that of QS (lower than -80 dBm). As shown in Fig.~\ref{fig:architecture}(b), in order to avoid the impact of IC-XT, the waveband that CSs occupy will not be used for QSs. Thus IC-XT is turned to out-band noise and can be eliminated by filters. Then SRS is the main impairment source to QKD, including SRS in MCF and that in SSMF. To alleviate the SRS in MCF, a dedicated core is utilized to transmit QSs since SRS generated from CSs has a stronger impact on QSs in the same core. In Fig. \ref{fig:architecture}(a), core-N is used to transmit QSs and SSs. In SSMF, upstream signals have a greater impact on QSs than downstream signals (explained in Sec.\ref{section:experiment}) and the wavelengths of upstream signals are required to set higher than those of corresponding QSs due to the smaller SRS coefficient \cite{eraerds2010quantum}.

\begin{figure}[h]
\centering\includegraphics[width=8cm]{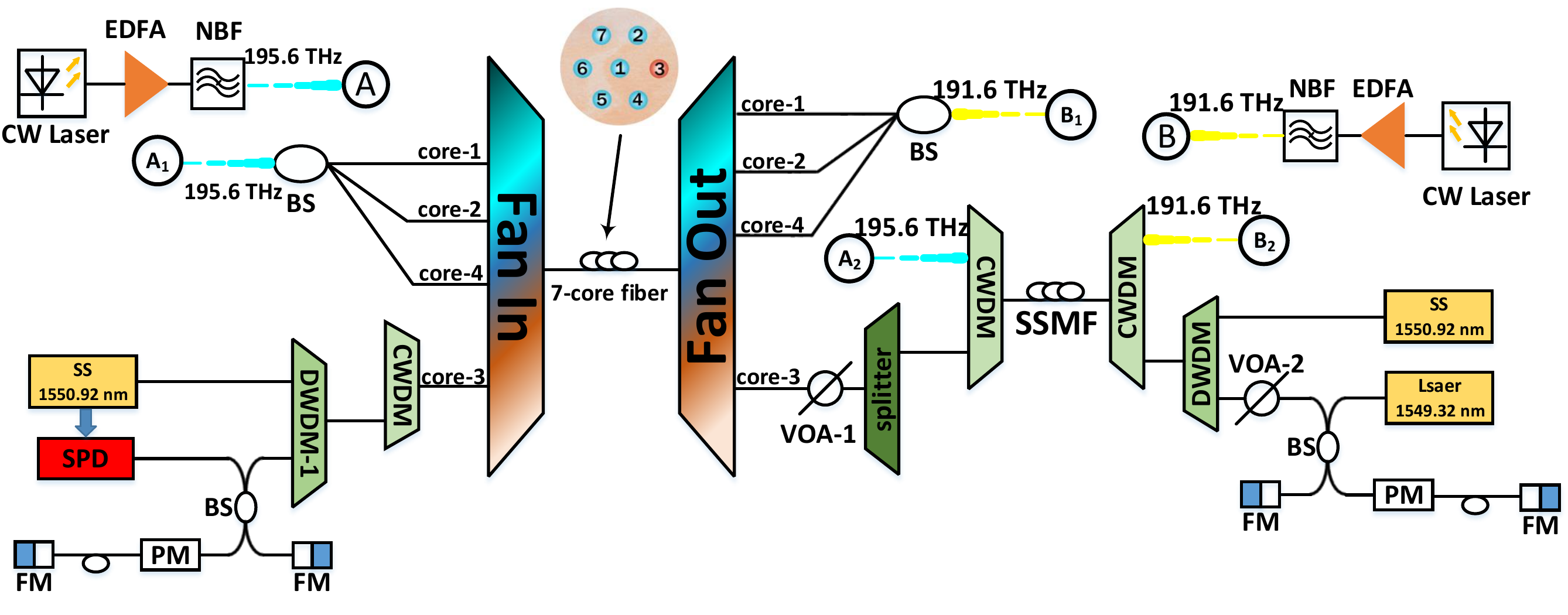}
\caption
{\label{fig:experiment}The experimental setup. ( SSMF: standard single mode fiber, DWDM: dense wavelength division multiplexing module, CWDM: coarse wavelength division multiplexing module, BS: beam splitter, SS: synchronization signal, VOA: variable optical attenuator, EDFA: erbium-doped fiber amplifier, NBF: narrow band filter, PM: phase modulator, FM: Faraday mirror, SPD: single photon detector.}
\end{figure}

The QKD receivers are placed in OLT to be shared by several ONUs since the single photon detectors (SPDs) in QKD receivers are often expensive and difficult to operate. Then we need to consider the multiplexing method of different QKD transmitters which are located in ONUs. Compared with the transmitting frequency of QKD transmitter, the receiving frequency of SPD is remarkably low and difficult to improve at present. If the QSs are based on TDM, the transmitting frequency of each ONU is forced to decrease. For example, the transmitting frequency of each ONU is 500 MHz under the case that two ONUs work based on TDM if the receiving frequency of SPD is 1 GHz. Thus the SKR of each ONU will decrease greatly when the number of users is large. Also, a splitter is required to couple the QSs to MCF. It will introduce 3 dB of loss each time the number of users is doubled. Hence, a network of 32 users has a minimum of 15 dB loss in the quantum channels, which will result in a significant reduction in SKR of QKD. Overall, TDM is not suitable for QSs since the proposed architecture is required to support a large amount of subscribers in the future. Instead, in a wavelength division multiplexing (WDM) based approach, the splitter is replaced by a wavelength multiplexer. It has less insertion loss than the splitter (a 32-channel wavelength multiplexer has a insertion loss of about 3 dB). However, the method that one QS occupy one wavelength will result in a great waste of wavelength resources since the wavelengths that QSs occupy can not be used for CSs. Also, different QKD receivers are required to receive QSs at different wavelengths. This will result in huge cost due to the expensive SPD. Above all, W-TDM is employed for QSs. This is realized by the CWDM module and splitters shown in Fig.~\ref{fig:architecture}(a). QKD transmitters connected to the same splitter are based on TDM while those connected to different ports of CWDM module are based on WDM. SSs are also necessray for QKD. They are set in the same waveband and are coupled into the MCF with the CWDM. QSs are transmitted from ONUs to OLT while SSs are transmitted from OLT to ONUs. By this means, QKD transmitters connected to the same splitter can be synchronized.

\section{Experimental setup and results}
\label{section:experiment}

\begin{figure}[h]
\centering\includegraphics[width=8cm]{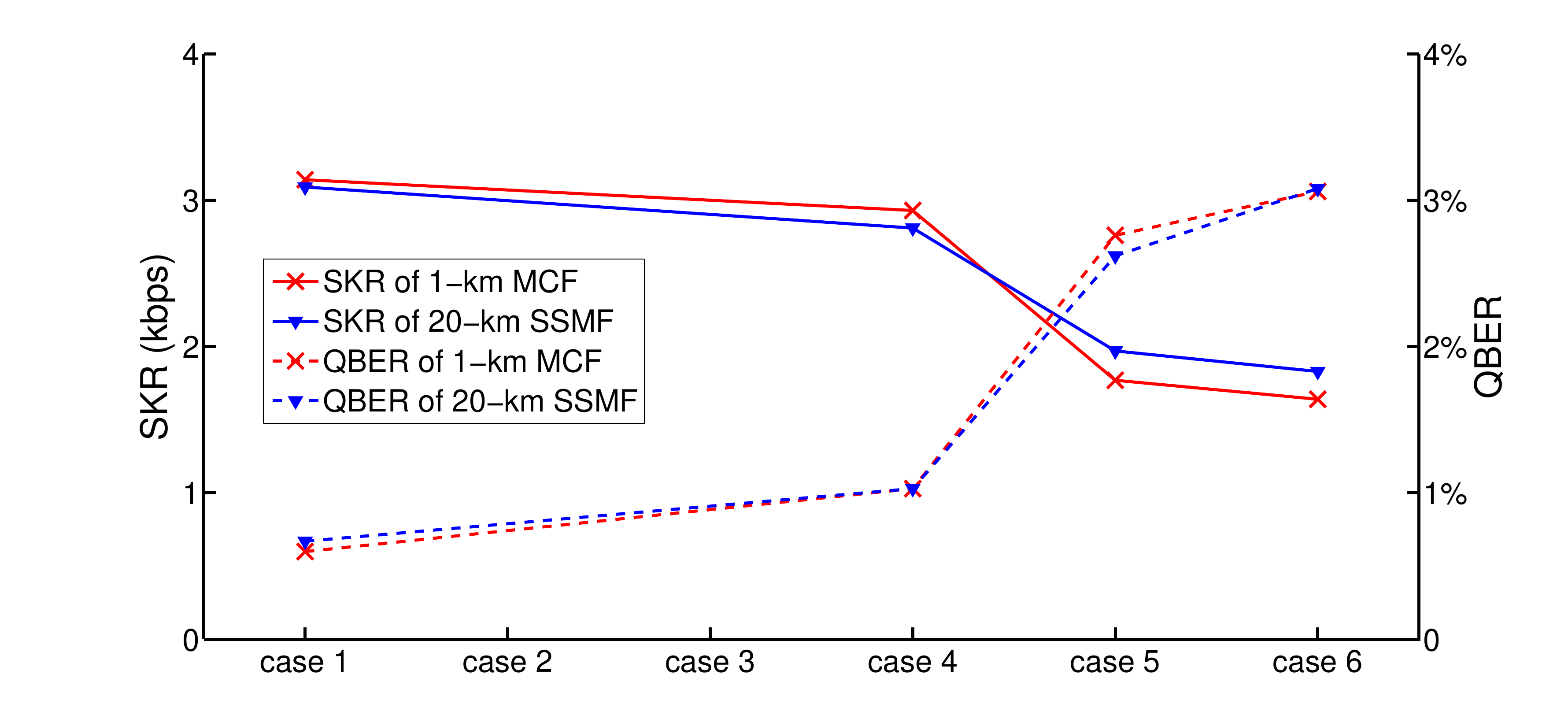}
\caption
{\label{fig:compareSSMF}Marks in case 1 represent the QBERs and the SKRs when no CS is transmitted in the system. Marks in case 4 represent the QBERs and the SKRs when the downstream signal is transmitted in drop fiber with the power of -10 dBm. Marks in case 5 represent the QBERs and the SKRs when the upstream signal is transmitted in drop fiber with the power of 0 dBm. Marks in case 6 represent the QBERs and the SKRs when the two kinds of signals mentioned above are transmitted in the system simultaneously.}
\end{figure}

The experiment setup is illustrated in Fig.~\ref{fig:experiment}. The QKD transmitter is located in the ONU. The QS is placed at 193.5 THz (1549.32 nm) and the SS is placed at 193.3 THz (1550.92 nm). Launch power of the SS is adjusted to maintain a received power of about -55 dBm. Thus SS will have no effcet on QS. BB84 phase coding scheme is used in our system combined with the decoy method. The averaged photon numbers of the signal state, decoy state, and vacuum state are chosen to be 0.6, 0.2, and 0, respectively \cite{Wang2005Beating, Ma2005Practical}. Alice launches the three types of states at a ratio of 14:1:1. The system operates at a frequency of 50 MHz. The entire QKD postprocessing is based on Ethernet, including the error correction with a low-density parity-check (LDPC) algorithm, error verification with a cyclic redundancy check. The detection efficiency of the SPD is about 8\% with an effective gating width of 1 ns and a dark count rate per gate of $10^{-6}$ in average. The passband of DWDM-1 in Fig.~\ref{fig:experiment} is about 150 GHz. The upstream signal (191.6 THz) and the downstream signal (195.6 THz) are sent by the continuous wave (CW) laser source. Narrow band filters (NBFs) are used to filter the amplified spontaneous emission (ASE) noise. The 7-core fiber used as feeder fiber in the experiment is 1 km. Core-3 is used to transmit QSs while other cores are used to transmit CSs. The SSMF in Fig.~\ref{fig:experiment} is used as drop fiber and the length is 1 km. $A$ is connected to $A_{1}$ or $A_{2}$ in Fig.~\ref{fig:experiment} if the downstream signal needs to be injected into MCF or SSMF, while $B$ is connected to $B_{1}$ or $B_{2}$ if the upstream signal needs to be injected into MCF or SSMF.

We want to evaluate the performance of the system in a more realistic scenario where the length of feeder fiber is about 20 km. Thus variable optical attenuator-1 (VOA-1) is set to 4.6 dB to simulate the loss of 20 km MCF (0.23 dB/km). Firstly, we would like to verify the rationality for this simulation. We replace the MCF (with Fan-in and Fan-out devices) and VOA-1 with a 20-km SSMF. Also an extra loss of 3.6 dB is introduced to make up for the loss of Fan-in and Fan-out when the 20-km SSMF is used. In the experiments, we maintain the receiving power of the upstream siganls and downstream signals higher than -15 dBm. Thus the power of downstream signal injected into the drop fiber is set to -10 dBm and the transmission power of upstream signal is set to 0 dBm. Splitter is not used in this comparative experiments and results are shown in Fig.~\ref{fig:compareSSMF}. Every point in Fig.~\ref{fig:compareSSMF} represents the average value of two-hour interval, which is the same with other Figs. in this paper. The marks in Fig.~\ref{fig:compareSSMF} indicate the experiment results, while the solid line and the dotted line are for explaining the trend of the experiment results and have no specific practical significance. As can be seen, the QBERs and SKRs for MCF and the 20-km SSMF are almost the same. Thus it is reasonable to simulate longer MCF with VOA-1 in the proposed architecture.

\begin{figure}[h]
\centering\includegraphics[width=8cm]{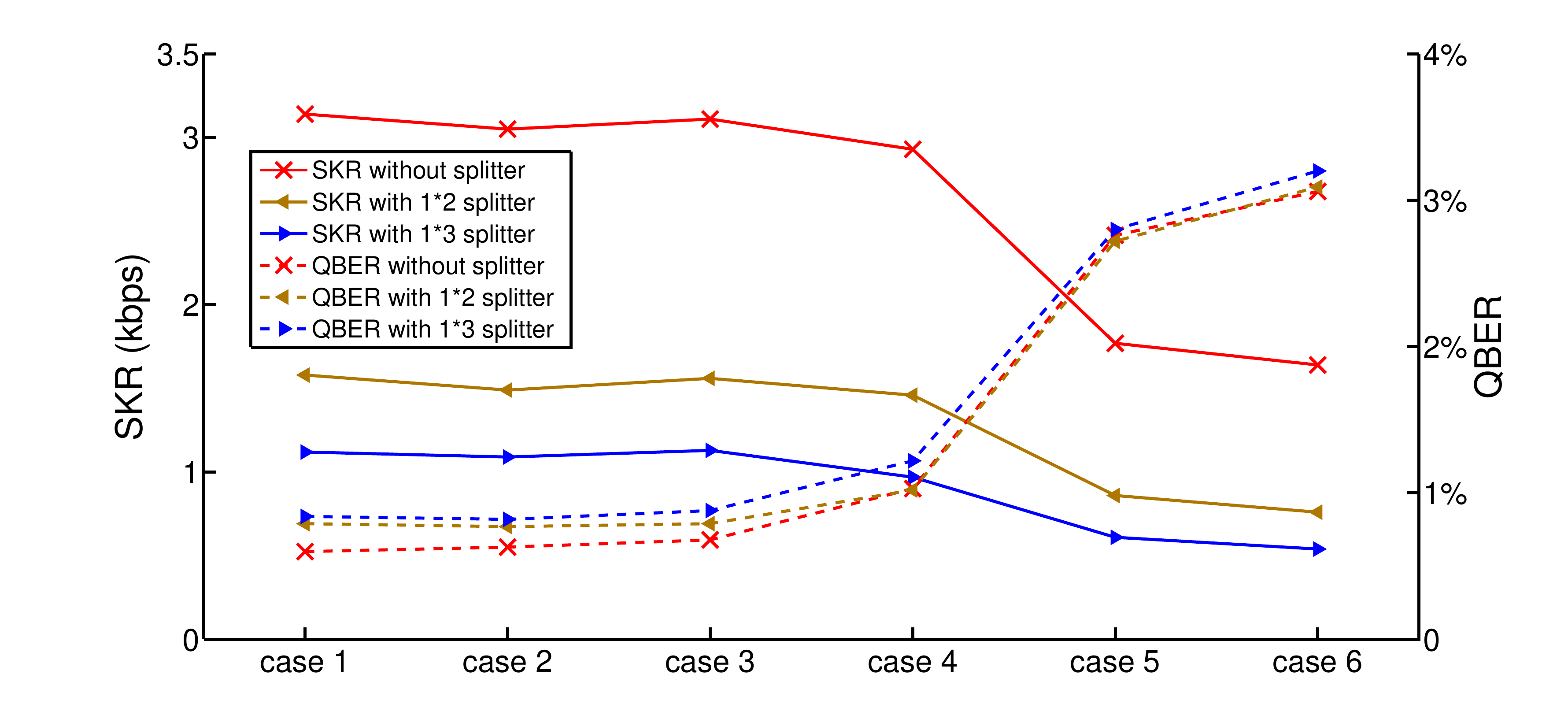}
\caption
{\label{fig:twofactors}Marks in case 1 represent the QBERs and the SKRs when no CS is transmitted in the system. Marks in case 2 represent the QBERs and the SKRs when the downstream signals are transmitted in core-1, -2, -4 of MCF. The power of downstream signals at each input port of the Fan-in is 20 dBm. Marks in case 3 represent the QBERs and the SKRs when the upstream signals are transmitted in core-1, -2, -4 of MCF. The power of upstream signals at each input port of the Fan-in is 20 dBm. Marks in case 4 represent the QBERs and the SKRs when the downstream signal is transmitted in drop fiber with the power of -10 dBm. Marks in case 5 represent the QBERs and the SKRs when the upstream signal is transmitted in drop fiber with the power of 0 dBm. Marks in case 6 represent the QBERs and the SKRs when the four kinds of signals mentioned above are transmitted in the system simultaneously.}
\end{figure}

The two main factors affecting the SKR of QKD are the channel loss and the QBER. The change of quantum channel loss is mainly rely on the change of the splitter. The main noise in the system is IC-XT and SRS. The CSs and QSs are assigned in different wavebands according to the proposed CWAS. Thus IC-XT is turned into out-band noise and can be eliminated by filters. Then the total QBER of QKD in the architecture can be expressed by
\begin{equation}
\begin{split}
\begin{aligned}
\label{eq:QBER}
QBER_{total}&=QBER_{MCF}+QBER_{SSMF}+QBER_{In}\\
&=QBER_{MCF-DS}+QBER_{MCF-US}\\
&+QBER_{SSMF-DS}+QBER_{SSMF-US}\\
&+QBER_{In},
\end{aligned}
\end{split}
\end{equation}
where $QBER_{total}$ is the total QBER of QKD, $QBER_{In}$ is the inherent QBER of the system, $QBER_{MCF}$ is the QBER caused in MCF, $QBER_{SSMF}$ is the QBER caused in SSMF, $QBER_{MCF-DS}$ is the QBER caused by downstream signals in MCF, $QBER_{MCF-US}$ is the QBER caused by upstream signals in MCF, $QBER_{SSMF-DS}$ is the QBER caused by downstream signal in SSMF, $QBER_{SSMF-US}$ is the QBER caused by upstream signal in SSMF. We evaluate the two factors in the experiments and the results are shown in Fig.~\ref{fig:twofactors}. Over the entire measurement, both SKR and QBER display very small fluctuations around their average values with a standard deviation of less than 0.39 kbps and 0.44\%, respectively. We only transmit classical signals in core-1, -2 and -4 since the impacts from the nearest neighbour cores are much larger than those from non-nearest neighbour cores. The QBERs in case 1-6 correspond to the items $QBER_{In}$, $QBER_{In}+QBER_{MCF-DS}$, $QBER_{In}+QBER_{MCF-US}$, $QBER_{In}+QBER_{SSMF-DS}$, $QBER_{In}+QBER_{SSMF-US}$ and $QBER_{total}$. As can be seen, the SKRs and QBERs in case 2 and 3 are almost the same as those in case 1. Thus the SRS in MCF can be ignored for QSs, which means the two terms $QBER_{MCF-DS}$ and $QBER_{MCF-US}$ can be ignored in Eq. (\ref{eq:QBER}). The QBER is dependent on the noise generated in SSMF. The upstream signal has a larger impact on QKD than downstream signal due to its higher power, which means $QBER_{SSMF-US}$ is the dominate term in Eq. (\ref{eq:QBER}). The SKR decreases dramatically with the increase of the splitting ratio due to the larger insertion loss. We have to emphasize that the transmitting frequency for each user is not changed in the experiments with 1*2 and 1*3 splitter. However, in practical application, the transmitting frequency for each user will decrease with the increase of splitting ratio. It also verifies that TDM is not suitable for quantum access networks with large amount of users due to the large insertion loss of splitter with high splitting ratio.


\begin{figure}[h]
\centering\includegraphics[width=8cm]{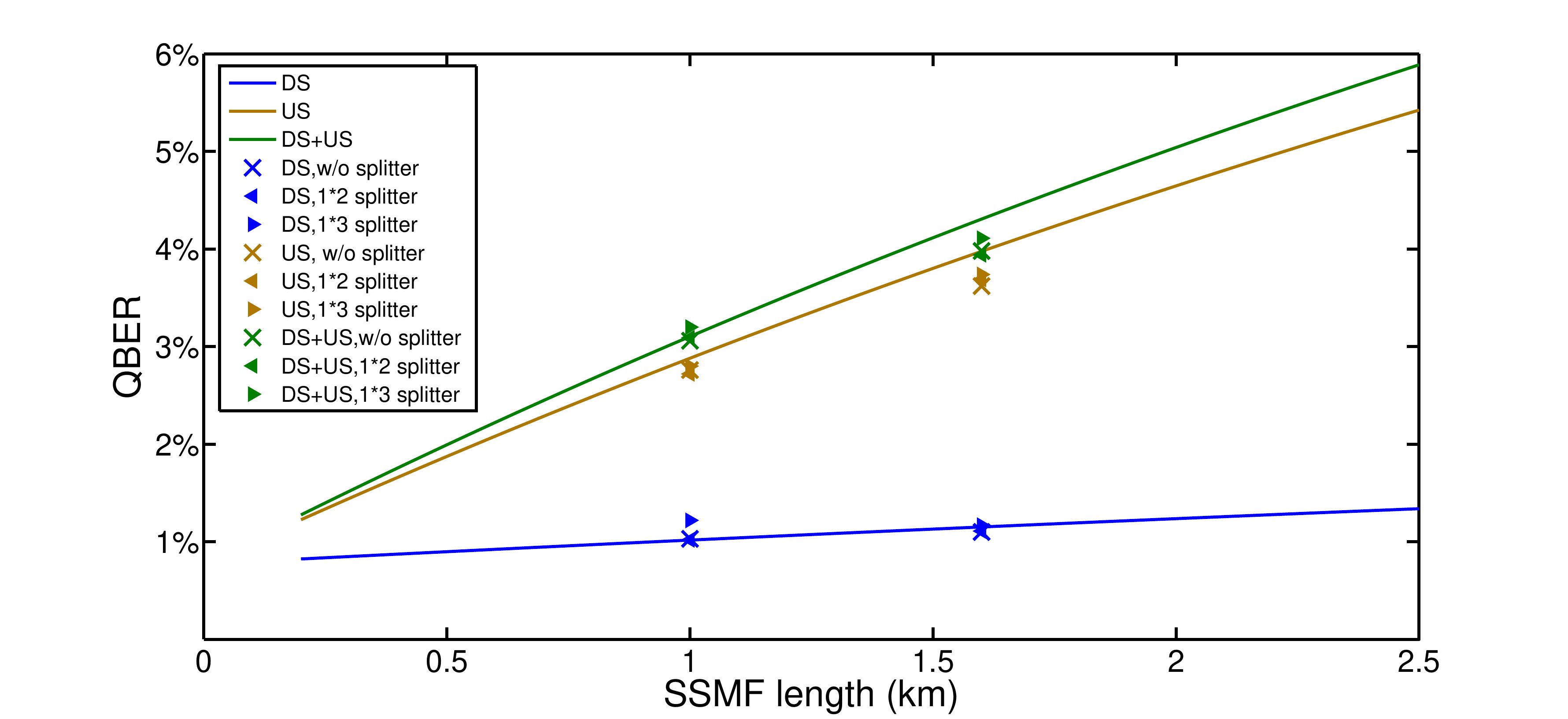}
\caption
{\label{fig:differentqber}The green solid line is the QBER when the downstream and upstream signals are transmitted in the SSMF simultaneously. The brown solid line is the QBER when the upstream signal of 0 dBm is transmitted in the SSMF.  The blue solid line is the QBER when the downstream signal of -10 dBm is transmitted in the SSMF. The marks are the experiment results.}
\end{figure}

The SRS generated in SSMF is the main impairment source to QKD. We evaluate the QBER caused by different signals separately with SSMF length. Another SSMF of 1.6 km is used as drop fiber. The theoretical simulation is performed based on the methods in \cite{Ma2005Practical} and \cite{PhysRevX.2.041010}. The results are shown in Fig.~\ref{fig:differentqber}. Upstream signal is the leading factor of QBER due to its higher power in SSMF. Thus the frequency of upstream signal is required to set lower than that of QS due to the smaller SRS coefficient \cite{eraerds2010quantum}.

\begin{figure}[h]
\centering\includegraphics[width=8cm]{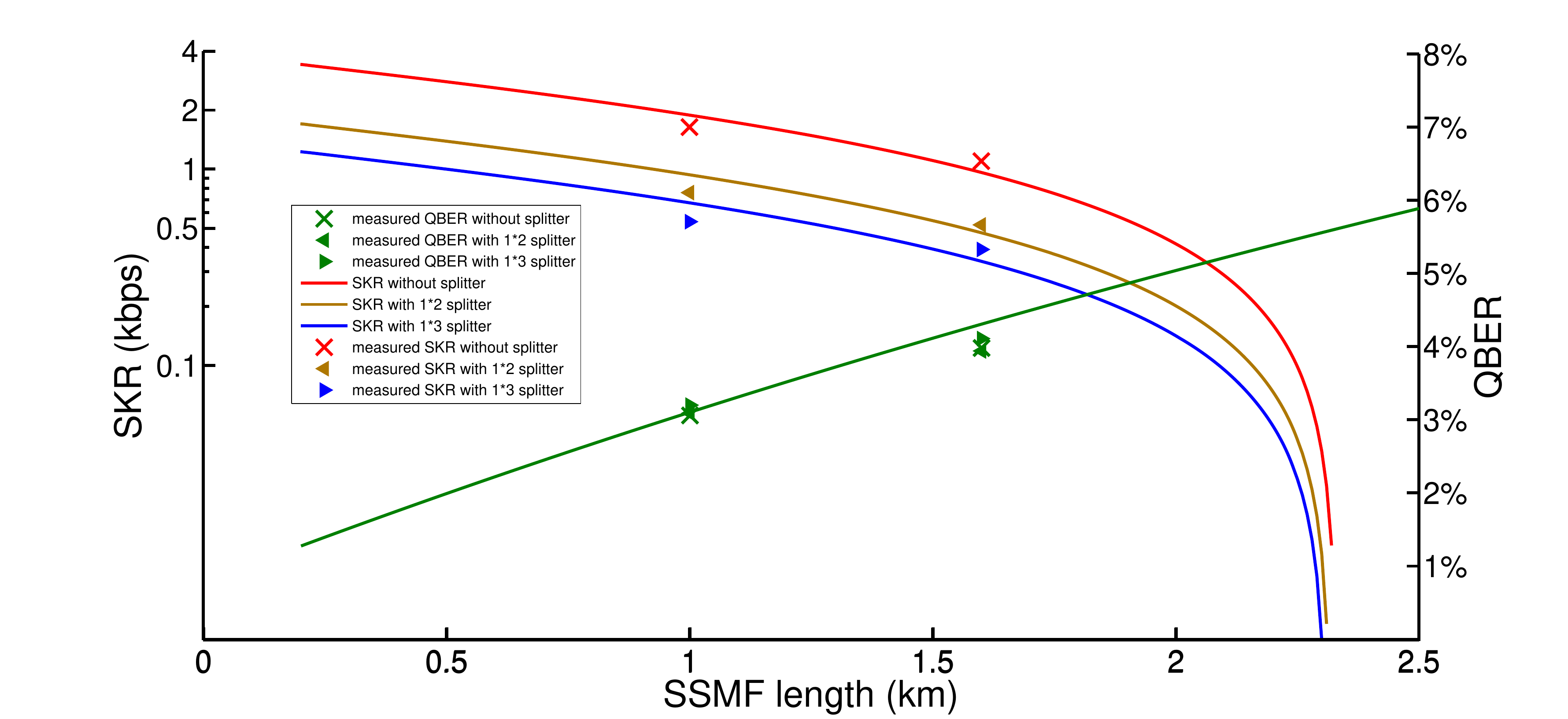}
\caption
{\label{fig:dysystem}SKR and the total QBER with the SSMF length based on our experimental setup.}
\end{figure}

As the QBER is mainly produced in SSMF, the length of SSMF is the main factor affecting system performance. We plot the SKR and the total QBER with the SSMF length in  Fig.~\ref{fig:dysystem}. As can be seen, the SKR decrease greatly with the increase of SSMF length and the maximum length that can be supported is less than 2.5 km. This is because no strict frequency or temporal filtering methods are used in the experiments (The passband of DWDM-1 is 150 GHz and the effective gating width of SPD is 1 ns).


\begin{figure}[h]
\centering\includegraphics[width=8cm]{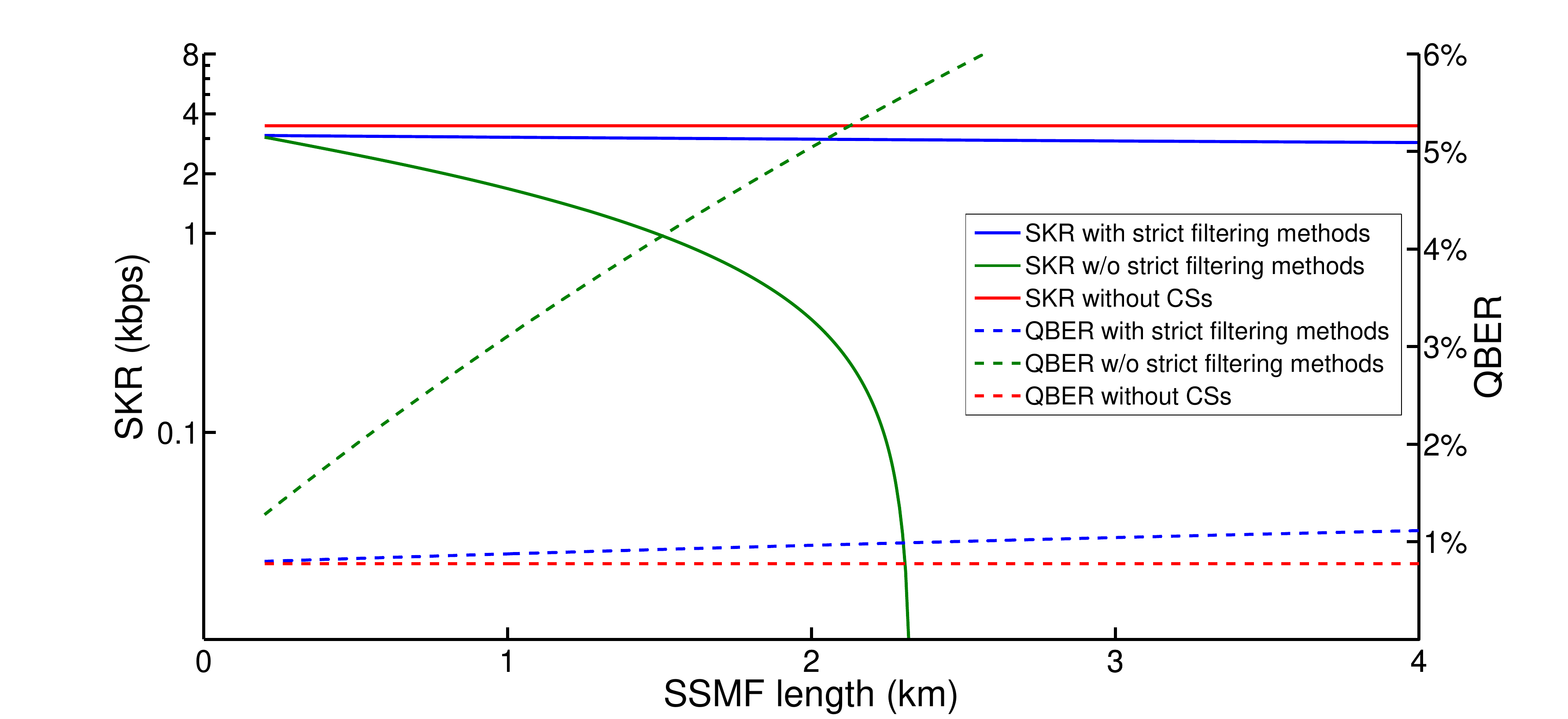}
\caption
{\label{fig:goodxiaoguo}SKR and QBER with the SSMF length based on our experimental setup. The passband of the filter is about 30 GHz with a insertion loss of 0.8 dB. The effective gating width of SPD is set to 0.18 ns.}
\end{figure}

We simulate the QBER and SKR when introducing more strict filtering methods. More specifically, an extra bandpass filter is used in the simulation and the effective gating width of SPD is set to 0.18 ns. The QBER and SKR with filtering methods are very close to those without CSs as shown in Fig.~\ref{fig:goodxiaoguo}. The discrepancies between the two QBERs and SKRs are about 0.18\% and 0.4 kbps when the length of SSMF is 2 km (typical length of drop fiber in access networks). The discrepancy between the SKRs is mainly due to the insertion loss of the filter.

\begin{figure}[h]
\centering\includegraphics[width=8cm]{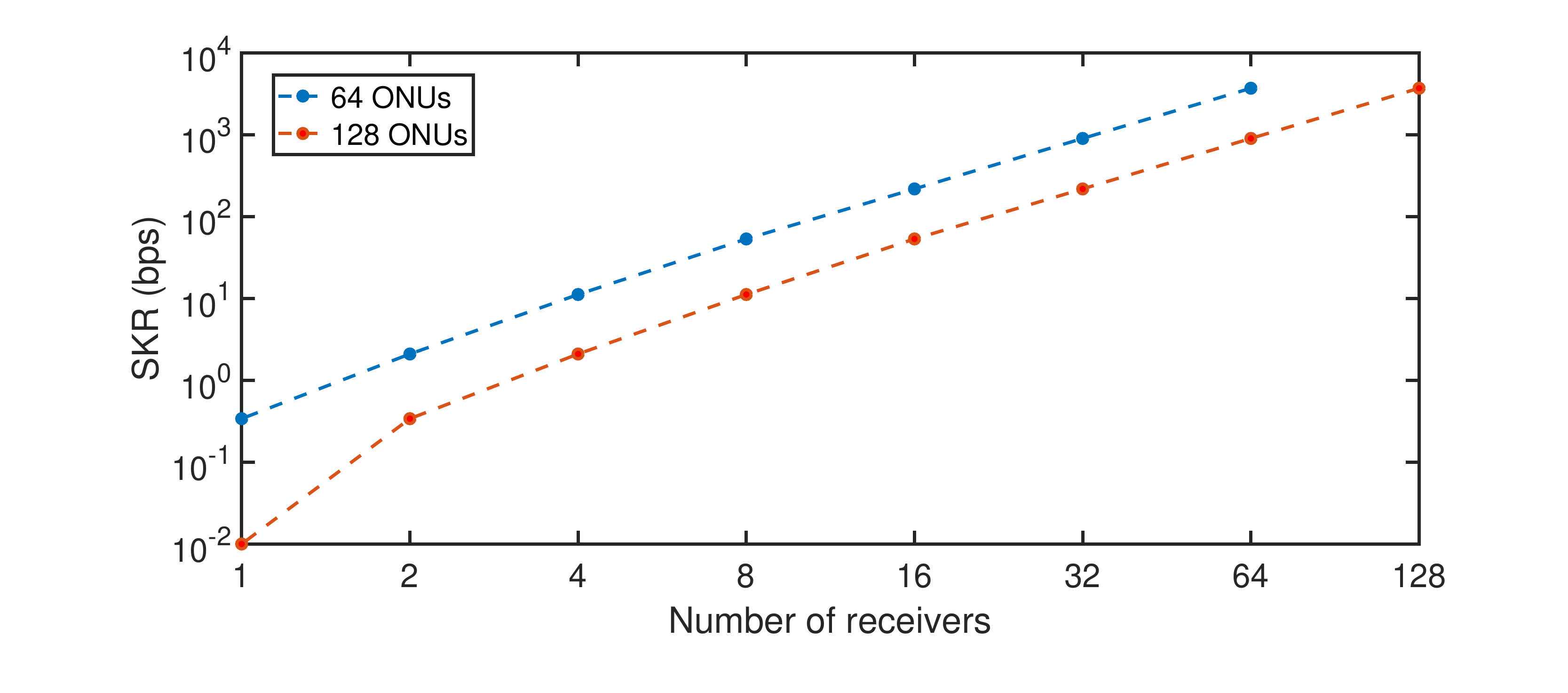}
\caption
{\label{fig:SPDnum}Relationship between SKR and the number of receivers based on the experiment system.}
\end{figure}

Finally, we simulate the relationship between the number of QKD receivers and SKR of each subscribers based on our system. The change in the number of receivers is realized by changing the splitting ratio of the splitter, for example, a $1*32$ beam splitter is required for 64 ONUs when two receivers are used in the network. The insertion loss of the splitters is set as 3.2 dB, 6.3 dB, 9.2 dB, 12.7 dB, 16.3 dB, 19.6 dB, and 22.8 dB for $1*2$, $1*4$, $1*8$, $1*16$, $1*32$, $1*64$, and $1*128$, respectively \cite{frohlich2015quantum}. We consider both the insertion loss of the splitter and the change of transmitting frequency in the simulation. As can be seen from Fig.~\ref{fig:SPDnum}, the more receivers are used in the network, the higher SKR can be achieved. In practical applications, we can choose the number of receivers reasonably according to the requirement for the SKR.


\section{Conclusion}
In this paper, we propose a MCF-based quantum access network which can support enormous subscribers. In the architecture, MCF is used as feeder fiber and SSMFs are used as drop fibers. Cost and SKR are two main concerns. In order to save deployment cost, QSs are integrated with CSs in both MCF and SSMFs. Also, the expensive QKD receivers are placed in the OLT, which will be shared by several QKD transmitters. To increase the SKR, a CWAS is proposed to alleviate the SRS and IC-XT noise from the intense CSs and QKD transmitters operate based on W-TDM. Finally, the architecture is verified experimentally and the experiment results are in good agreement with the theoretical analysis. The SKR of each ONU can reach 1.64 kbps for a transmission distance of 20 km, even without strict frequency or temporal filtering methods. Our results pave the way to extending the applications of QKD to last mile communications in the future.


%





\ifCLASSOPTIONcaptionsoff
  \newpage
\fi



%


\bibliographystyle{IEEEtran}
\bibliography{IEEEabrv,bare_jrnl}

%








\end{document}